\documentclass[groupedaddress,showpacs,showkeys,amssymb,eqsecnum,aps]{revtex4}
\usepackage{lmodern}
\usepackage[]{graphicx}
\usepackage[]{graphics}
\usepackage{amsmath}
\usepackage{epsf}
\usepackage{slashed}

\usepackage{color}                   
\usepackage{verbatim}
\usepackage[mathscr]{euscript}
\usepackage{hyperref}

\newcommand{\be}{\begin{equation}}

\newcommand{\ee}{\end{equation}}

\newcommand{\tr}{{\rm Tr\,}}

\begin{document}

%\begin{comment}
 \author{F. T. Brandt}  
 \email{fbrandt@usp.br}
 \affiliation{Instituto de F\'{\i}sica, Universidade de S\~ao Paulo, S\~ao Paulo, SP 05508-090, Brazil}

\author{J. Frenkel}
\email{jfrenkel@if.usp.br}
\affiliation{Instituto de F\'{\i}sica, Universidade de S\~ao Paulo, S\~ao Paulo, SP 05508-090, Brazil}

 \author{D. G. C. McKeon}
 \email{dgmckeo2@uwo.ca}
 \affiliation{
 Department of Applied Mathematics, The University of Western Ontario, London, Ontario N6A 5B7, Canada}
 \affiliation{Department of Mathematics and Computer Science, Algoma University, Sault Ste.~Marie, Ontario P6A 2G4, Canada}
 
% \end{comment}

\title{Feynman diagrams in terms of on-shell propagators}

\date{\today}

\begin{abstract}
It is shown that the usual expression for a Feynman diagram in terms of the Feynman %--Stueckelberg
propagator $\Delta_F(x-y)$ can be replaced by an equivalent expression involving the positive-energy on-shell propagator $\Delta^+(x-y)$, supplemented by appropriate functions associated with time-ordering.  When this alternate way of expressing a Feynman diagram is Fourier transformed into momentum space, the momentum associated with each function $\Delta^+(x-y)$ is on-shell, and is only conserved at each vertex if an energy is attributed to the contributions of the time-ordering functions. The resulting expression is analogous to what Kadyshevsky had obtained by deriving an alternate expansion for the $S$--matrix. A detailed explanation of how this alternate expansion is derived is given, and it is shown how it provides a straightforward way of determining the imaginary part of a Feynman diagram, which makes it useful when using unitarity methods for computing a Feynman diagram. By considering a number of specific Feynman diagrams in self-interacting scalar models and in QED, we show how this alternate approach can be related to the old perturbation theory and can simplify direct calculations of Feynman diagrams.
\end{abstract}

\pacs{11.15.Bt,11.55.-m} %;11.55.−m}
%\pacs{11.10Kk,04.60Kz}
%PACS No.: 11.15.-q \\                                              
%KEY WORDS: gauge theories,  first order,  perturbation theory                                             
\keywords{Perturbation Theory, Unitarity Methods}

\maketitle

\section{Introduction}\label{sec1} 
A key ingredient of the $S$--matrix for a scattering process in quantum field theory is the vacuum expectation value of the time-ordered product of fields \cite{PeskinSchroeder,weinberg2005quantum,Schwartz2013}. The original approach to computing these ``Green's functions'' lacked manifest covariance and was cumbersome to use \cite{heitler1984quantum}. With the advent of Feynman perturbation theory \cite{feynman1962quantum}, perturbative calculation became streamlined and manifest covariance was retained, leading to almost universal adoption of this approach \cite{kaiser2009drawing}.

The covariant perturbative approach to computing $S$-matrix elements that originated with Feynman involves the use of the causal (or Stueckelberg-Feynman) propagator \cite{Stueckelberg:1941rg,feynman1962quantum}
%If we were to consider a scalar theory in six dimensions with cubic interactions, this propagator is
\be\label{e1}
\Delta_F(x-y) = \theta(x^0-y^0) \Delta^+(x-y) + \theta(y^0-x^0) \Delta^-(x-y),
\ee
where $\Delta^\pm(x-y)$ are positive (negative) energy on-shell functions
\be\label{e2}
\Delta^\pm(x-y) = \int\frac{d^4 k}{(2\pi)^3 i}e^{-ik\cdot (x-y)}\theta(\pm k^0)
\delta(k^2-m^2) = \Delta^\mp(y-x)
\ee
We first restrict our attention to self-interacting scalar fields. Any Feynman loop may be expressed as a linear combination of scalar loops, by using an integral reduction procedure \cite{tHooft:1978jhc}.

Representing the step function $\theta$ in terms of an integral
\be\label{e3}
\theta(x^0-y^0) = \int_{-\infty}^{\infty} \frac{d\tau}{2\pi i}
\frac{e^{i\tau (x^0-y^0)}}{\tau-i\epsilon}
\ee
we see from Eqs. \eqref{e1}, \eqref{e2} and \eqref{e3} that
\be\label{e4}
\Delta_F(x-y) =\int\frac{d^4 k}{(2\pi)^4}\frac{e^{-ik\cdot(x-y)}}{k^2-m^2+i\epsilon} =\Delta_F(y-x) .
\ee
(A detailed discussion of $\Delta_F$ is given in Ref. \cite{pauli2000selected}.) %xxxx

We can express a Feynman diagram in the form %\cite{r6}
\be\label{e5}
I_{n,m}(y_1,y_2,\cdots,y_m) = \int dx_1 \cdots dx_n 
\Delta_F(y_1-x_1)\cdots\Delta_F(y_m-x_m)
\left[\Delta_F(x_i-x_j)\cdots\Delta_F(x_k-x_l)\right],
\ee
where $m$ is the number of vertices associated with an external propagator and $n$ is the total number of vertices, located at points $x_1$, $x_2$, $\cdots$, $x_n$. With a cubic interaction, there are $(3n-m)/2$ propagators $\Delta_F(x_i-x_j)$ within the square brackets in Eq. \eqref{e5}. Since $\Delta_F$ is manifestly covariant, systematic ways of computing Feynman diagrams can be developed (see, for example, Refs. \cite{PeskinSchroeder,weinberg2005quantum,Schwartz2013}).

An alternate way of perturbatively expanding the $S$--matrix was developed by Kadyshevsky \cite{Kadyshevsky:1967rs} (see also \cite{Fuda:1986zg}). It involves use of the on-shell propagator $\Delta^+(x-y)$ of Eq. \eqref{e2} as well as the time ordering function $\theta$ of Eq. \eqref{e3}. In this paper, we first show in Sec. \ref{sec2} that the expansion of Kadyshevsky can be derived from the Feynman expansion.

In Secs. \ref{sec3} and \ref{sec4} we discuss a number of applications of using the functions $\Delta^+$ and $\theta$ in a perturbative expansion of $S$. One useful feature is that the imaginary part of a Feynman diagram can be found immediately, and is equivalent to the cutting rules of Cutkosky \cite{Cutkosky:1960sp} (see Appendix \ref{appA}). Using the imaginary part of a Feynman diagram in conjunction with a dispersion relation makes it possible to compute the Feynman diagram without calculating the usual Feynman integral over an off-shell loop momentum \cite{Elvang:2015rqa}.
This approach also allows for a simple and direct integration over the internal loop energies, which enables to establish useful connections between on-shell tree amplitudes and loop integrals \cite{Feynman:1963axKlauder:1972lsv,Berger:2009zb,Brandhuber:2005kd,Rodrigo:2008fp,Bierenbaum:2010cy,Caron-Huot:2010fvq,Brandt:2021nseBrandt:2021nev}.
In Sec. \ref{sec5} we extend the analysis to fermionic fields and give an application to the vacuum polarization in QED, while in Sec. \ref{sec6} we apply this method to the calculation of the amplitude for the decay process $\pi_0\rightarrow 2\gamma$
We conclude the paper with a brief discussion in Sec. \ref{sec7}.

%zzzzzzzzzzzzzzzzzzzzzzzzz
%\cite{schweber2011introduction,weinberg2005quantum,heitler1984quantum,feynman1962quantum,kaiser2009drawing,Stueckelberg:1941rg,pauli2000selected,Kadyshevsky:1967rs,Cutkosky:1960sp,Elvang:2015rqa,Gribov:2000nhz,Berestetskii:1982qgu,Broadhurst:1987ei,vanNeerven:1985xr,tHooft:1972fi,Bollini:1972ui,Feynman:1963ax,Klauder:1972lsv,Caron-Huot:2010fvq,tHooft:1973wag,Veltman:1994wz,Isgur:1972qs}

\section{Use of the functions $\Delta^+$ and $\theta$}\label{sec2}  % in place of $\Delta_F$}
\begin{figure}[t!]
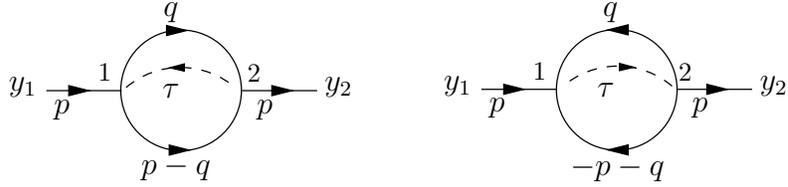

\input fig1aRev.pspdftex $\qquad$ $\qquad$
\input fig1b.pspdftex
\caption{Graphical representation of the two terms in Eqs. \eqref{e8} and \eqref{e10}.}
\label{fig1}
\end{figure}

We begin by considering the one-loop, two-point function when there is a cubic interaction in $D=d+1$ space-time dimensions. In this case, Eq. \eqref{e5} becomes
%\cite{schweber2011introduction,weinberg2005quantum}  %\cite{Caron-Huot:2010fvq} %xxxxxxx
\be\label{e6} 
i I^{(1)}_{2,2}(y_1,y_2) = \int d^Dx_1 d^Dx_2 
\Delta_F(y_1-x_1) \Delta^2_F(x_1-x_2) \Delta_F(x_2-y_2).
\ee
Since
\begin{subequations}\label{e7}
  \be\label{e7a}
\theta(z) \theta(-z) = 0 
\ee
and
  \be\label{e7b}
\theta(z) \theta(z) = 1 ,
  \ee 
\end{subequations}
Eq. \eqref{e1} reduces Eq. \eqref{e6} to
\be\label{e8}
i I^{(1)}_{2,2}(y_1,y_2) = \int d^Dx_1  d^Dx_2 
\Delta_F(y_1-x_1) \left[
\theta(x_1^0-x_2^0)  {\Delta^{+}_F}^2(x_1-x_2) +
\theta(x_2^0-x_1^0)  {\Delta^{+}_F}^2(x_2-x_1) 
\right]
  \Delta_F(x_2-y_2),
\ee
which now involves only the functions $\theta$ and $\Delta^+$. Since
\be\label{e9}
\theta(\pm k_0)\delta(k^2-m^2) = \frac{\delta(k_0\mp\omega_k)}{2\omega_k};\;\;\;\;\; \left(\omega_k\equiv\sqrt{|\vec k|^2+m^2}\right),
\ee
Eqs. \eqref{e2}, \eqref{e3} and \eqref{e4} now show that 
\begin{eqnarray}\label{e10}
i I^{(1)}_{2,2}(y_1,y_2) &=& \frac{i}{4}\int \frac{d^D p}{(2\pi)^D} 
\frac{e^{-p\cdot(y_1-y_2)}}{(p^2-m^2+i\epsilon)^2}  \nonumber \\
& & \int \frac{d^d q}{(2\pi)^d}\int_{-\infty}^{\infty}\frac{d\tau}{\tau-i\epsilon} 
\left[
\frac{\delta(p^0-\omega_q-\omega_{p-q}+\tau)}{\omega_q\omega_{p-q}}+
\frac{\delta(p^0+\omega_q+\omega_{-p-q}-\tau)}{\omega_q\omega_{-p-q}}
\right] .
\end{eqnarray}
upon integrating over $x_1$ and $x_2$ in Eq. \eqref{e8}. The two terms in Eqs. \eqref{e8} and \eqref{e10}
can be represented graphically in Fig \ref{fig1}.

\begin{figure}[b!]
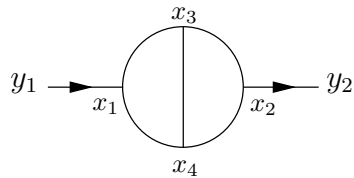

\input fig2.pspdftex
  \caption{Diagram which corresponds to the integral in Eq. \ref{e13}.}
\label{fig2}
\end{figure}

The argument used to obtain Eq. \eqref{e10} from Eq. \eqref{e6} will now be applied to the Feynman diagram of Fig. \ref{fig2}.

In general, if there are $n$ vertices at $x_1^\mu$, $x_2^\mu$, $\dots$, $x_n^\mu$, their time component can be ordered in $n!$ ways   
\be\label{e11}
x^{0}_{(1)}\ge x^{0}_{(2)}\cdots\ge x^{0}_{(n)}.
\ee
For the diagram of Fig. \ref{fig2}, let us consider the time ordering
\be\label{e12}
x^{0}_{3}\ge x^{0}_{4}\ge x^{0}_{1}\ge x^{0}_{2}.
\ee
The Feynman diagram of Fig \ref{fig2} corresponds to the integral
\begin{eqnarray}\label{e13}
I^{(2)}_{4,2}(y_1,y_2) &=& \int d^Dx_1d^Dx_2d^Dx_3d^Dx_4\Delta_F(y_1-x_1)
\nonumber \\ &&
\left[
                \Delta_F(x_1-x_3)\Delta_F(x_1-x_4)\Delta_F(x_3-x_4)\Delta_F(x_3-x_2)
                \Delta_F(x_4-x_2)\right]\Delta_F(x_2-y_2).
\end{eqnarray}
The propagators $\Delta_F$ with the square brackets of Eq. \eqref{e13} can result in a product of five $\theta(x^0_i-x^0_j)\equiv\theta_{ij}$. The time ordering of Eq. \eqref{e12} is contained within the product
\be\label{e14}
\theta_{34}\theta_{31}\theta_{32}\theta_{41}\theta_{42}.
\ee
The product of Eq. \eqref{e14} also contains the time ordering
\be\label{e15}
x^{0}_{3}\ge x^{0}_{4}\ge x^{0}_{2}\ge x^{0}_{1}.
\ee
Two of the $4!$ time orderings of the vertices in Fig. \ref{fig2} are given explicitly in Eqs. \eqref{e12} and \eqref{e15}; these correspond to a contribution 
\begin{eqnarray}\label{e16}
J_{4,2}^{(2)}(y_1,y_2) &=& \int d^D x_1 d^D x_2 d^D x_3 d^D x_4 \Delta_F(y_1-x_1)
\left[\theta_{34}\theta_{41}\theta_{12}+\theta_{34}\theta_{42}\theta_{21}\right]
\nonumber \\ &&
\left[\Delta^+_{34}\Delta^+_{41}\Delta^+_{42}\Delta^+_{31}\Delta^+_{32}\right]
\Delta_F(x_2-y_2)\;\;\; (\Delta^+_{ij}\equiv\Delta^+(x_i-x_j) ) .
\end{eqnarray}
In general, a Feynman diagram with $n$ vertices, $m$ of which are external, has $(3n-m)/2$ internal lines and $(n-m)/2 + 1$ loops when there is  cubic self-interaction. Each distinctive time-ordering of the vertices according to Eq. \eqref{e11} results in a factor of
\be\label{e17}
\theta_{(1)(2)}\theta_{(2)(3)}\cdots\theta_{(n-1)(n)}.
\ee
This is multiplied by $(3n-m)/2$ factors of $\Delta^+(x_i-x_j)$ in place of the factor $\Delta_F(x_i-x_j)$ in the original expression of Eq. \eqref{e5} with $x_i^0\ge x_j^0$. Upon using Eqs. \eqref{e2} and \eqref{e3} and doing the integrals over $x_1^\mu,\cdots, x_n^\mu$, we find that we have the graphical rules of Fig. \ref{fig3}, with conservation of spatial components of momentum at each vertex, and with the sum of the temporal components of momentum entering a vertex equaling the sum of the $\tau_i$ entering that vertex. This leads to a factor of $(2\pi)^D$ at each vertex. These rules are consistent with those derived by Kadyshevsky \cite{Kadyshevsky:1967rs}.

\begin{figure}[t!]
\input fig3a.pspdftex $\qquad$ $\qquad$ 
\input fig3bD.pspdftex $\qquad$ $\qquad$ 
\input fig3cD.pspdftex 
\caption{Graphical rules obtained upon using Eqs. \eqref{e2} and \eqref{e3} and doing the integrals over $x_1^\mu,\cdots, x_n^\mu$. There is conservation of spatial components of momentum at each vertex; the sum of the temporal components of momentum entering a vertex is equal to the sum of the $\tau_i$ entering that vertex.} 
\label{fig3}
\end{figure}

\begin{figure}[b!]
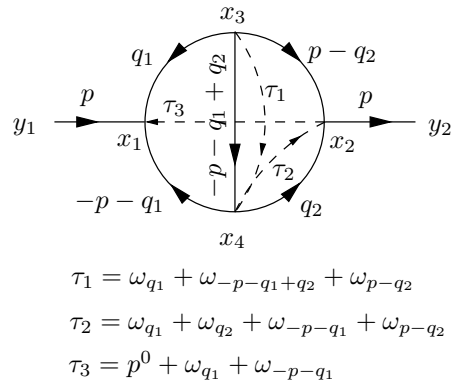

\input fig4.pspdftex 
  \caption{Graphical representation of Eq. \eqref{e18}.} 
\label{fig4}
\end{figure}

If we were to apply these rules to the second term in Eq. \eqref{e16} to the Feynman diagram of Fig. \ref{fig2} (eq. \eqref{e15}), we obtain
\begin{eqnarray}\label{e18}
J^{(2)}_{4,2}(y_1,y_2) &=& \int \frac{d^D p}{(2\pi)^D}\frac{e^{-ip\cdot(y_1-y_2)}}{(p^2-m^2+i\epsilon)^2}
\int \frac{d^d q_1}{(2\pi)^d}\frac{d^d q_2}{(2\pi)^d}
\frac{1}{2\omega_{-p-q_1+q_2}}\frac{1}{2\omega_{q_2}}\frac{1}{2\omega_{p-q_2}}
\frac{1}{2\omega_{q_1}}\frac{1}{2\omega_{-p-q_1}}
\nonumber \\ &&
\left[
\frac{1}{\omega_{q_1}+\omega_{-p-q_1+q_2}+\omega_{p-q_2}-i\epsilon}
\frac{1}{\omega_{q_1}+\omega_{q_2}+\omega_{-p-q_1}+\omega_{p-q_2}-i\epsilon}
\frac{1}{p^0+\omega_{q_1}+\omega_{-p-q_2}-i\epsilon}
\right] . 
\end{eqnarray}
This is represented diagrammatically in Fig. \ref{fig4}.

This is associated with the time ordering of Eq. \eqref{e15}; the remaining 23 contribution associated with the Feynman diagram of Fig. \ref{fig2} can be found in a similar fashion.

%%% Insert A

\begin{figure}[t!]
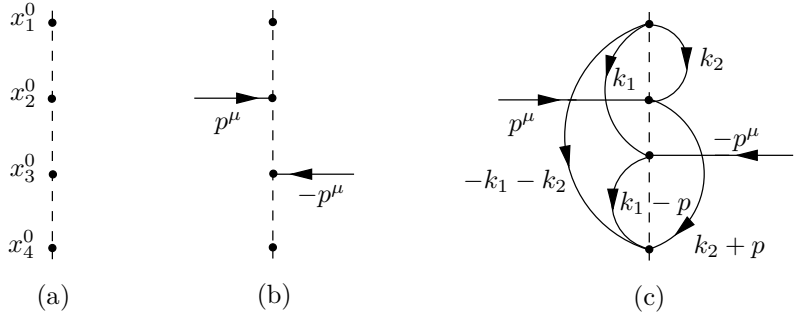

\input fig5nc.pspdftex 
  \caption{Stages in constructing a two-loop, two-point diagram.}
\label{fig5n}
\end{figure}

The rules that have been outlined can be recast in a different way. If one were to consider all contributions with $n$ vertices, $m$ of which are external, occurring at $x^\mu_1$, $x^\mu_2$ $\dots$ $x^\mu_n$, with 
$x^0_1 \ge x^0_2 \dots \ge x^0_n$, then this can be represented by a vertical line as in Fig. \ref{fig5n}a for $n=4$.

To $m$ of these vertices, an external leg with momentum $p_i^\mu$ ($i=1,\dots,m$) can be attached as in Fig. \ref{fig5n}b for $n=4$, $m=2$ ($\sum_i p_i^\mu =0$). All vertices are now connected so that each vertex is associated with three lines as in Fig. \ref{fig5n}c. The momentum in each connecting line has momentum flowing from $x^\mu_i$  to $x^\mu_j$ with $x_i^0>x_j^0$ and spatial momentum conserved at each vertex.

\begin{figure}[b!]
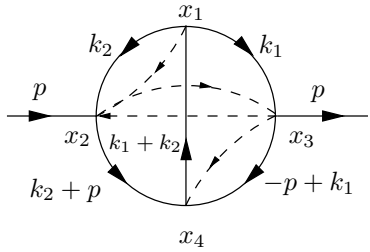

\input fig6n.pspdftex 
  \caption{Alternative representation of the diagram of Fig. \ref{fig5n}c.}
\label{fig6n}
\end{figure}

Each dotted line connecting $x_i^0$ to $x^0_{i+1}$ ($i=1,2\dots,n-1$) has $\tau_i$ chosen so that the temporal momentum is conserved at each vertex. In Fig. \ref{fig5n}c this means that we have
\begin{subequations}\label{e214}
\be\label{e214a}
\tau_1 = \omega_{k_2}+\omega_{-k_1-k_2} + \omega_{k_1},
\ee
\be\label{e214b}
\tau_2 = \omega_{-k_1-k_2}+\omega_{k_1} + \omega_{k_2+p} -p^0
\ee
and
\be\label{e214c}
\tau_3 = \omega_{k_1-p}+\omega_{-k_1-k_2} + \omega_{k_2+p}.
\ee  
\end{subequations}
The diagram of Fig. \ref{fig5n}c can also be represented in Fig. \ref{fig6n}.

\begin{figure}[t!]
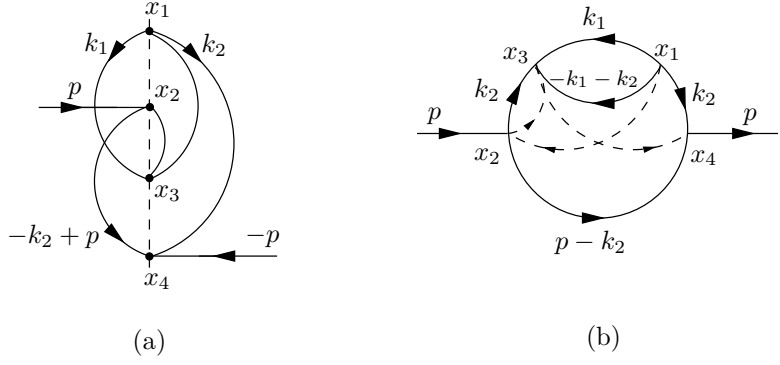

\input fig7n.pspdftex 
\caption{Graphical representations of Eq. \eqref{e216}.}
\label{fig7n}
\end{figure}

The integral associated with Figs. \ref{fig5n} and \ref{fig6n} is
\begin{eqnarray}\label{e215}
&&\int\frac{d^D p}{(2\pi)^D} \int\frac{d^d k_1}{(2\pi)^d} \frac{d^dk_2}{(2\pi)^d}
\frac{1}{(2\omega_{k_1})(2\omega_{k_2})(2\omega_{-k_1-k_2})(2\omega_{k_2+p})(2\omega_{k_1-p})}
\frac{e^{-ip\cdot(y_2-y_1)}}{(p^2-m^2+i\epsilon)^2}
\nonumber \\ &&
\frac{1}{\left(\omega_{k_2}+\omega_{-k_1-k_2}+\omega_{k_1}-i\epsilon \right)
%\frac{1}{
         \left(\omega_{-k_1-k_2}+\omega_{k_1}+\omega_{k_2+p}-p^0-i\epsilon \right)
%\frac{1}{
         \left(\omega_{k_1-p}+\omega_{-k_1-k_2}+\omega_{k_2+p}-i\epsilon\right)}.
\end{eqnarray}
Similarly, the diagram of Fig \ref{fig7n}a, or alternatively, Fig. \ref{fig7n}b, is associated to the integral
\begin{eqnarray}\label{e216}
&&\int\frac{d^D p}{(2\pi)^D} \int\frac{d^d k_1}{(2\pi)^d} \frac{d^d k_2}{(2\pi)^d}
\frac{1}{(2\omega_{k_1}) (2\omega_{k_2})^2(2\omega_{-k_1-k_2})(2\omega_{-k_2+p})} 
\frac{e^{-ip\cdot(y_2-y_1)}}{(p^2-m^2+i\epsilon)^2}
\nonumber \\ &&
\frac{1}{\left(\omega_{k_1}+\omega_{-k_1-k_2}+\omega_{k_2}-i\epsilon\right)
%\frac{1}{
         \left(\omega_{-k_1-k_2}+\omega_{k_1}+\omega_{-k_2+p}+p^0-i\epsilon\right)
%\frac{1}{
         \left(\omega_{k_2}+\omega_{-k_2+p}-p^0-i\epsilon\right)}.
\end{eqnarray}

\section{Application to loop calculations}\label{sec3} %  Application of using $\Delta^+$ and $\theta$}
Having demonstrated how to express a Feynman diagram in terms of the on-shell propagator $\Delta^+$, we will consider how this alternate approach can be used. We first note that the imaginary contribution to a Feynman diagram must come from the step function $\theta$.

Since
\be\label{e19}
\frac{1}{A\pm i\epsilon}= {\cal P}\frac{1}{A}\mp i\pi\delta(A)
\ee
we find, for example, that in Eq. \eqref{e10}
\begin{eqnarray}\label{e20}
\Im I(p^2) &=& \Im \int\frac{d^d q}{(2\pi)^d}\int_{-\infty}^{\infty}
\frac{d\tau}{\tau-i\epsilon}
               \left[
\frac{1}{\omega_q\omega_{p-q}}\delta(p^0-\omega_q-\omega_{p-q}+\tau)+
\frac{1}{\omega_q\omega_{-p-q}}\delta(p^0+\omega_q+\omega_{-p-q}-\tau)
\right]
\nonumber \\ 
           &=& \pi \int\frac{d^d q}{(2\pi)^d}\frac{\delta(p^0+2\omega_q)+\delta(p^0-2\omega_q)}{\omega_q^2},
\end{eqnarray}
if we consider the frame of reference in which $\vec p =0$, so that $p^0=\pm\sqrt{p^2}$. Since in $d$ dimensions \cite{tHooft:1972fiBollini:1972ui}
\be\label{e21}
\int d^d q f(|\vec q|) = \frac{2\pi^{d/2}}{\Gamma(d/2)}\int_0^\infty dq |\vec q|^{d-1}f(|\vec q|)
\ee
and as
\be\label{e22}
\int_{-\infty}^\infty dx f(x) \delta(g(x))=
\sum_i \frac{f(a_i)}{|g^\prime(a_i)|}\;\;\;\; (g(a_i)=0),
\ee
%\be\label{e23}
%\Im I(p^2) = \frac{16\pi^{3} (p^2/4-m^2)^{3/2}}
%{3(2\pi)^5\sqrt{p^2} } \theta(p^2-4 m^2).
%\ee
%
\be\label{e23}
\Im I(p^2) = \frac{4\pi}{(2\pi)^d}\frac{\pi^{d/2}}{\Gamma(d/2)} \frac{1}{\sqrt{p^2}}
%\frac{16\pi^{3}
\left(\frac{p^2}{4}-m^2\right)^{d/2-1}
%{3(2\pi)^5\sqrt{p^2} }
    \theta(p^2-4 m^2).
\ee

This approach to computing the imaginary part of a Feynman diagram is equivalent to using ``cut'' Feynman propagators \cite{Cutkosky:1960sp}.
%The vertices associated with the dashed line arising from each choice of time ordering are associated with the circled vertices in the ``largest time'' approach of Refs. \cite{r14} to using unitarity to evaluate a Feynman diagram.  
At one-loop order, Feynman essentially used the expansion outlined here to demonstrate the existence of ghost contributions in Yang-Mills theory and gravity \cite{Feynman:1963axKlauder:1972lsv}.
%his approach has been extended beyond one-loop order \cite{Feynman:1963axKlauder:1972lsv,Berger:2009zb,Brandhuber:2005kd,Rodrigo:2008fp,Bierenbaum:2010cy,Caron-Huot:2010fvq,Brandt:2021nseBrandt:2021nev}.
%
%\cite{Caron-Huot:2010fvq}.

One can now consider a direct computation of the integrals in Eq. \eqref{e10}, which involve, in $d$ dimensions, the function
\be\label{e29}
I(p^2) = \int\frac{d^dq}{(2\pi)^d}\int_{-\infty}^\infty\frac{d\tau}{\tau-i\epsilon}
\left[\frac{1}{\omega_q\omega_{p-q}}\delta(p^0-\omega_q-\omega_{p-q}+\tau)+
    \frac{1}{\omega_q\omega_{-p-q}}\delta(p^0+\omega_q+\omega_{-p-q}-\tau) 
\right] .
\ee
This is a Lorentz invariant quantity proportional to the self-energy function, which is more easily evaluated in the reference frame $\vec p = 0$. Setting $m^2=0$ and first integrating over $\tau$, we obtain the result
\be\label{e33}
I(p^2) = \frac{8}{(16\pi)^{d/2}} \frac{\Gamma(d/2-1/2)\Gamma(3/2-d/2)}{\Gamma(d/2)}
(-p^2)^{d/2-3/2} .
\ee

In Eq. \eqref{e29} we could also set $\vec p = 0$ and then integrate over $\vec q$ in $d$ dimensions before integrating over $\tau$. This procedure leaves us with
\begin{eqnarray}\label{e35}
  I(p^2) &=& \int_{-\infty}^\infty \frac{d\tau}{\tau-i\epsilon}
  \left( \frac{2\pi^{d/2}}{(2\pi)^d\Gamma(d/2)}  \right)
             \int_0^\infty \frac{dq q^{d-1}}{q^2+m^2}
             \left[
\delta(p^0 - 2 \sqrt{q^2+m^2}+\tau)+\delta(p^0 + 2 \sqrt{q^2+m^2}-\tau)
             \right]
\nonumber \\ &=&
                 \frac{8}{(2\pi)^d}
\frac{\pi^{d/2}}{\Gamma(d/2)}\int_{2 m}^\infty\frac{(\tau^2/4-m^2)^{d/2-1}}{\tau^2-p^2-i\epsilon} d\tau.
\end{eqnarray}
The integral over $q$ is well defined and the divergence only
arises from the integral over $\tau$.
This is consistent with Ref. \cite{Kadyshevsky:1967rs}.
The above integral has precisely the form of a dispersion relation. This may be explicitly evaluated, for $m^2=0$, when it leads to the result given in Eq.  \eqref{e33}. For $m\neq 0$, it may be expressed in terms of the Gauss hypergeometric function ${}_2F{_1}(1,(3-d)/2;3/2;p^2/4 m^2)$.

\begin{figure}[t!]
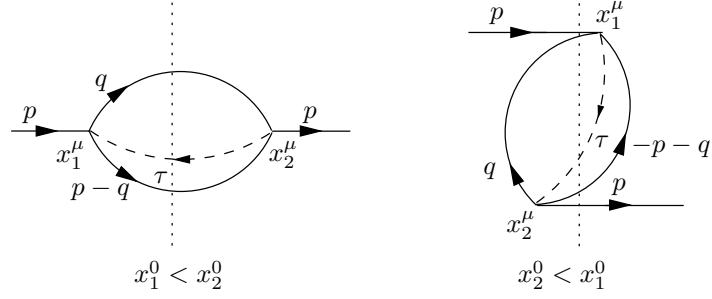

\input fig5aRev.pspdftex $\qquad$ $\qquad$
\input fig5bRev.pspdftex 
\caption{Graphical representation of the two terms in Eq. \eqref{e29}.}
\label{fig5}
\end{figure}

It is interesting to note how Eq. \eqref{e29} is related to what one encounters when using the old perturbation theory (OPT) \cite{heitler1984quantum}. In this approach, if $S=1-2\pi i T$, then one solves for the matrix elements of $T$ iteratively,
\be\label{e36}
T_{\beta\alpha} = V_{\beta\alpha}+
\int d\gamma\frac{V_{\beta\gamma}V_{\gamma\alpha}}{E_\alpha-E_\gamma+i\epsilon}+
\int d\gamma d\gamma^\prime
\frac{V_{\beta\gamma}V_{\gamma\gamma^\prime}V_{\gamma^\prime\alpha}}
{(E_\alpha-E_\gamma+i\epsilon)(E_\alpha-E_\gamma^\prime+i\epsilon)} + \dots ,
\ee
where $V$ is the interaction part of the Hamiltonian \cite{heitler1984quantum}.
If we integrate Eq. \eqref{e29} over $\tau$, we get
\be\label{e37}
I(p^2) = \int \frac{d^d q}{(2\pi)^d}
\left[
\frac{1}{\omega_q\omega_{p-q}}\frac{1}{-p^0+\omega_q+\omega_{p-q}-i\epsilon}+
\frac{1}{\omega_q\omega_{-p-q}}\frac{1}{p^0+\omega_q+\omega_{-p-q}-i\epsilon}
\right],
\ee
which corresponds to %the two terms in Eq. \eqref{e37} correspond to
the diagrams of Fig \ref{fig5}.
The vertical cuts in Fig \ref{fig5} cut lines associated with intermediate states in the second term in the sums of Eq. \eqref{e36}. The factors $1/(\omega_q\omega_{p-q})$ and $1/(\omega_q\omega_{-p-q})$ in Eq. \eqref{e37} are associated with the wave function of the virtual intermediate states.

%The two-loop self-energy ``sunrise'' diagram of Fig. \ref{fig6} that occurs in four dimensions with a quartic interaction for a scalar field now will be considered.

\section{The sunrise diagram}\label{sec4}
\begin{figure}[b!]
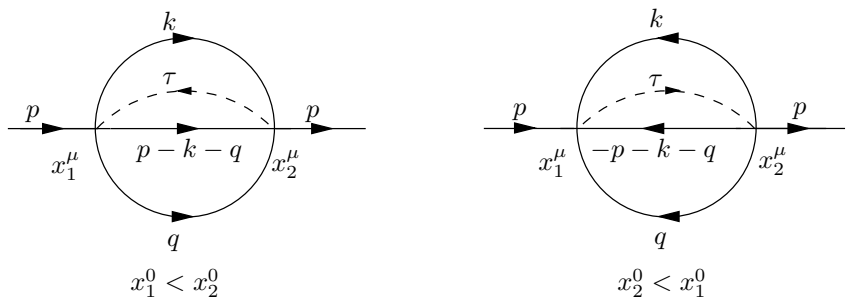

\input fig6aRev.pspdftex $\qquad$ $\qquad$
\input fig6bRev.pspdftex 
\caption{Two-loop sunrise self-energy diagrams with a quartic interaction, corresponding
to the integral in Eq. \eqref{e39a}.}
\label{fig6}
\end{figure}

As a workable application at two loops, let us now consider the self-energy diagram of Fig. \ref{fig6} that occurs in four dimensions with a quartic interaction for a scalar field.

The two diagrams associated with the time orderings indicated in Fig. \ref{fig6} result in the Feynman integral
\be\label{e38}
I^{(2)}_{2,2}(y_1,y_2)=\int d^4x_1 d^4x_2 \Delta_F(y_1-x_1)[\Delta_F(x_1-x_2)]^3 \Delta_F(x_2-y_2)
\ee
being expressed as
\begin{eqnarray}\label{e39a}
I^{(2)}_{2,2}(y_1,y_2)
  &=&\frac{1}{8}\int\frac{d^4p}{(2\pi)^4}\frac{e^{-ip\cdot(y_1-y_2)}}{(p^2-m^2+i\epsilon)}
\int\frac{d^3 q}{(2\pi)^3}\frac{d^3 k}{(2\pi)^3} \int_{-\infty}^\infty \frac{d\tau}{\tau-i\epsilon}
\nonumber \\ &&                
\left[\frac{\delta(p^0-\omega_k-\omega_q-\omega_{p-k-q}+\tau)}{\omega_k\omega_q\omega_{p-k-q}}+
     \frac{\delta(p^0+\omega_k+\omega_q+\omega_{-p-k-q}-\tau)}{\omega_k\omega_q\omega_{-p-k-q}}
                \right]
\end{eqnarray}
upon using the rules obtained earlier.

Performing the $\tau$-integration, leads to the following expression for the scalar self-energy at two-loops
\begin{equation}\label{e39}
\Pi(p) = \frac{\lambda^2}{3!}
\int\frac{d^3 q}{(2\pi)^3}\frac{d^3 k}{(2\pi)^3} 
\frac{1}{2\omega_k}\frac{1}{2\omega_q}\frac{1}{2\omega_{p+k+q}}
% \nonumber \\ &&
                 \left[
\frac{1}{\omega_k+\omega_q+\omega_{p+k+q}-p^0-i\epsilon} +
\frac{1}{\omega_k+\omega_q+\omega_{p+k+q}+p^0-i\epsilon}
                \right] ,
\end{equation}
where we made the transformations $\vec k\rightarrow -\vec k$ and $\vec q\rightarrow -\vec q$ in the first term. Note that the second term in Eq. \eqref{e39} can be obtained from the first term by making $p^0\rightarrow -p^0$.

Let us now evaluate the imaginary part of the above equation, which corresponds, by unitarity, to the decay rate of a scalar particle into three on-shell particles with energies $\omega_{k}$, $\omega_{q}$ and $\omega_{p+k+q}$
\begin{equation}\label{e40}
\Im \Pi(p) = \frac{\lambda^2}{3!} \pi
\int\frac{d^3 q}{(2\pi)^3}\frac{d^3 k}{(2\pi)^3} 
\frac{1}{2\omega_k}\frac{1}{2\omega_q}\frac{1}{2\omega_{p+k+q}}
% \nonumber \\ &&
                 \left[
\delta(\omega_k+\omega_q+\omega_{p+k+q}-p^0) +
\delta(\omega_k+\omega_q+\omega_{p+k+q}+p^0) 
                \right] .
\end{equation}
Let us take, for definiteness, $p^0 > 0$, so that only the first delta function will contribute to the imaginary part of the self-energy. For simplicity, we will consider the massless case with $m=0$.
By Lorentz invariance, since $p^\mu$ must be timelike in order for this process to occur, we may do the calculation in the frame where $\vec p=0$. In this case $\omega_k=|\vec k|\equiv k$,
$\omega_q=|\vec q|\equiv q$ and $\omega_{k+q}=\sqrt{k^2+q^2+2 k q x}$, where $x$ is the cosine of the angle between $\vec k$ and $\vec q$. Thus, we must evaluate the integral over this angle which involves 
\be\label{e41}
\int_{-1}^{1} dx \delta[k+q+\sqrt{k^2+q^2+2 k q x}-p^0]=
{\theta(p^0-k-q)}\frac{p^0-k-q}{k q},
\ee
where we have used the relation \eqref{e22}, together with the fact that the argument of the $\delta$-function vanishes at the point
\be\label{e42}
x_0   =   1+ \frac{{p^0}^2 -2 p^0 (k+q)}{2 k q}.
\ee
It turns out that the requirement that $-1< x_0 < 1$, leads to the condition $k+q+|k-q|<p^0< 2 (k+q)$.
Using these relations and performing the remaining angular integrations in Eq. \eqref{e40}, we obtain
\begin{equation}\label{e43}
\Im \Pi(p) = \frac{\lambda^2}{3!} \frac{\pi}{(2\pi)^4}\frac{1}{4}
\int_0^\infty dk \int_0^\infty dq
\theta(p^0-k-q-|k-q|)\theta(2(k+q)-p^0).
\end{equation}
These integrals may be evaluated by using the change of variables $k+q=Q_+$ and $k-q=Q_-$. Then, a straightforward integration over the variables $Q_+$ and $Q_-$ leads to the following result 
\be\label{e44}
\Im \Pi(p) = \frac{\lambda^2}{3!} \frac{\pi}{(4\pi)^4}\frac{1}{2} p^2,
\ee
where we have set ${(p^0)}^2\rightarrow p^2$ by Lorentz invariance.

The above result was obtained in four space-time dimensions. In $D$-dimensions, the corresponding result can be written in the form
\be\label{e45}
\Im \Pi(p) = \frac{\lambda^2}{3!} \frac{\pi}{(4\pi)^D}
\frac{\Gamma^3(D/2-1)}{\Gamma(3D/2-3)}\frac{1}{\Gamma(D-2)}
(p^2)^{D-3},
\ee
which reduces to Eq. \eqref{e44} for $D=4$.

With the help of this result for the imaginary part of the sunrise graph, one can write a dispersion relation for the self-energy $\Pi(p)$
\be\label{e46}
\Pi(p) = \frac{1}{\pi} \int_{4 m^2}^\infty ds \frac{\Im \Pi(s)}{s-p^2-i\epsilon}.
\ee
Using the expression \eqref{e45} and rescaling $s=p^2 x$, we obtain in the massless case that
\be\label{e47}
\Pi(p) = \frac{\lambda^2}{3!} \frac{1}{(4\pi)^D}
\frac{\Gamma^3(D/2-1)}{\Gamma(3D/2-3)\Gamma(D-2)}
(-p^2)^{D-3}\int_0^\infty dx\frac{x^{D-3}}{x-1-i\epsilon}.
\ee
This integral can be performed using the formula 3.194 in \cite{grads}, which leads to the following result 
\be\label{e48}
\Pi(p) = \frac{\lambda^2}{3!} \frac{1}{(4\pi)^D}
\frac{\Gamma(3-D)\Gamma^3(D/2-1)}{\Gamma(3D/2-3)}
(-p^2)^{D-3},
\ee
where $-p^2\rightarrow-(1+i\epsilon) p^2$. The above expression agrees with that obtained in the massless case by using conventional Feynman propagators, as expected (see for example Eq. (C4) of \cite{Brandt:2021nseBrandt:2021nev}).
But the present approach is more convenient for clarifying the way that the unitarity of the $S$--matrix arises from the intermediate states. The corresponding result obtained for a non-zero mass is more involved and can be expressed in terms of elliptic integrals \cite{Delbourgo:2003ziDavydychev:2003cw}.

\section{Spinor fields}\label{sec5}
The use of on-shell propagators is also possible when dealing with spinor fields. The Feynman propagator, much like that for a scalar field in Eq. \eqref{e1} is given by
\begin{eqnarray}\label{51}
S_F(x) &=& \theta(x^0) S^+(x) + \theta(-x^0) S^-(x),
\nonumber \\ &=& \left(i\gamma\cdot\partial+m\right) \Delta_F(x),
\end{eqnarray}
where
\begin{eqnarray}\label{52}
S^{\pm}(x) &=& \left(i\gamma\cdot\partial+m\right)   \Delta^{\pm}(x)
\nonumber \\ &=& \int\frac{d^3 k}{(2\pi)^3 i} \frac{e^{-i(\pm\omega_k x^0 - \vec k\cdot\vec x)}}{2\omega_k}
(\pm\omega_k\gamma^0 - \vec k\cdot\vec\gamma+m).
\end{eqnarray}
(Only if $m=0$ does $S^\pm(x)=-S^\mp(-x)$.)

The vacuum polarization is given by
\begin{eqnarray}\label{53}
\Pi_{\mu\nu}(y_1,y_2) &=& \int d^4 x_1 d^4 x_2 D_{F\;\mu\lambda}(y_1-x_1)D_{F\;\nu\sigma}(y_2-x_2)
\nonumber \\ &&
 (-i e)^2 \, \tr \left[\gamma^\lambda S_F(x_1-x_2) \gamma^\sigma S_F(x_2-x_1) \right],
\end{eqnarray}
where
\be\label{54}
D_{F\;\lambda\sigma}(x) = -i \eta_{\lambda\sigma} \int\frac{d^4 p}{(2\pi)^4}
\frac{e^{-ip\cdot x}}{p^2+i\epsilon}.
\ee
Using Eqs. \eqref{e7} we find that
\be\label{55}
\gamma^\mu S_F(x)\gamma^\nu S_F(-x)=\theta(x^0)\, \gamma^\mu S^+(x)\gamma^\nu S^-(-x)+
                                   \theta(-x^0)\, \gamma^\mu S^-(x)\gamma^\nu S^+(-x) .
\ee
Following the same steps that lead from Eq. \eqref{e8} to \eqref{e29}, we find that
\begin{eqnarray}\label{56}
\Pi_{\mu\nu}(y_1,y_2) &=& \frac{-i e^2}{4} \int\frac{d^4p}{(2\pi)^4}\frac{e^{-ip\cdot(y_1-y_2)}}{(p^2+i\epsilon)^2}
                          \int\frac{d^3k}{(2\pi)^3}\frac{1}{\omega_{p+k}\omega_k}
                          \nonumber \\ &&
\tr\left[  \frac{\gamma_\mu(\omega_k\gamma^0+\vec k\cdot\vec\gamma+m)
     \gamma_\nu(-\omega_{p+k}\gamma^0+(\vec p+\vec k)\cdot\vec\gamma+m)}{-p^0+\omega_k+\omega_{p+k}-i\epsilon}\right.
\nonumber \\
                        && \;\;\;\;\; + \left. \frac{\gamma_\mu(-\omega_k\gamma^0+\vec k\cdot\vec\gamma+m)
     \gamma_\nu(\omega_{p+k}\gamma^0+(\vec p+\vec k)\cdot\vec\gamma+m)}{p^0+\omega_k+\omega_{p+k}-i\epsilon}
                                          \right].
\end{eqnarray}
Using the well known trace formulae, it follows that the vacuum polarization may be written in momentum space as
\begin{eqnarray}\label{57}
  \Pi_{\mu\nu}(p) &=& e^2\int\frac{d^3 k}{(2\pi)^3}\frac{1}{\omega_k\omega_{p+k}}\left(\frac{1}{p^0+\omega_k+\omega_{p+k}-i\epsilon}
                                            +\frac{1}{-p^0+\omega_k+\omega_{p+k}-i\epsilon}\right)
  \nonumber\\ &&\left[
m^2\eta_{\mu\nu}-\omega_k\omega_{p+k}\left(2\eta_{\mu 0}\eta_{\nu 0}  - \eta_{\mu \nu}\right)
+k^i(p+k)^j\left(\eta_{\mu i}\eta_{\nu j}+\eta_{\mu j}\eta_{\nu i}-\eta_{\mu \nu} \eta_{ij}\right)
                 \right].
\end{eqnarray}
%Since this is a Lorentz covariant tensor, it
This expression may be conveniently evaluated in the reference frame in which $p^\mu=(p^0,\vec 0)$. In this frame, it is easy to verify that $\Pi_{\mu\nu}$ satisfies the Ward identity $p^\mu\Pi_{\mu\nu}(p)=0$ at the integrand level (in any dimensions). %Thus
Since $p^\mu \Pi_{\mu\nu}(p)$ is a four-vector, this must hold in any frame so that $\Pi_{\mu\nu}$ should have the transverse form
\be\label{58}
\Pi_{\mu\nu}(p) = (\eta_{\mu\nu}p^2 - p_\mu p_\nu) \Pi(p^2).
\ee
It is interesting to note that this approach naturally preserves gauge invariance.
The imaginary part of the function $\Pi(p^2)$ may be easily evaluated from Eq. \eqref{57} to give, in the above reference frame where $p^2 = {(p^0)}^2$, 
\be\label{59}
\Im \Pi(p^2) = \frac{e^2}{4\pi}\frac{1}{3}\left(1+\frac{2 m^2}{p^2}\right)\sqrt{1-\frac{4 m^2}{p^2}}\theta(p^2-4 m^2).
\ee
Using this expression, one can write for $\Pi(p^2)$ a dispersion relation of the form given in Eq. \eqref{e46}. The integration can be performed and leads to the result
% as the one obtained by employing the usual Feynman rules
\be\label{510}
\Pi(p^2) - \Pi(0) = -\frac{e^2}{2\pi^2}\int_0^1 dx \, x(1-x) \ln\left[1-\frac{p^2}{m^2} x(1-x)\right],
\ee
which agrees with that obtained by using the covariant Feynman rules to compute the photon self-energy
\cite{PeskinSchroeder,weinberg2005quantum,Schwartz2013}.

\section{Pion decay to photons}\label{sec6}
\begin{figure}[b!]
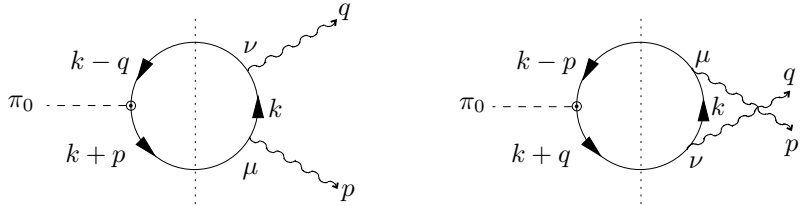

\input fig10a.pspdftex $\qquad$ $\qquad$
\input fig10b.pspdftex
\caption{Feynman diagrams contributing to the amplitude for the pion decay.}
\label{fig10n}
\end{figure}

This process takes place through an anomaly in the divergence of the axial-vector current \cite{AdlerBellJackiw}. In QCD, it provides evidence that the number of quark colors should be equal to 3, in agreement with the experimental data \cite{PeskinSchroeder,weinberg2005quantum,Schwartz2013}. Here, we will calculate the amplitude for the decay $\pi^0\rightarrow 2\gamma$ in QED, by using the present approach. The one-loop graphs for this process are shown in Fig. \ref{fig10n}, where the coupling of the pion to the fermions is $g_\pi\gamma^5$.
The sum of the diagrams gives
\be\label{e61}
\Gamma^{\mu\nu} = - g_\pi e^2 \int\frac{d^4 k}{(2\pi)^4}\tr\left[
\gamma^\nu S_F(k) \gamma^\mu S_F(k+p)\gamma^5 S_F(k-q)+ (\mu\leftrightarrow\nu, p\leftrightarrow q)
 \right].
\ee

In order to evaluate these contributions, one can use for the spinor propagators the form given in Eqs.
\eqref{51} and \eqref{52}. However, a simplification occurs if we first calculate the trace over the $\gamma$-matrices. To this end, we note that the trace of $\gamma^5$ with any number, less than 4, of $\gamma$-matrices vanishes. The trace of $\gamma^5$ with four $\gamma$-matrices, is %given by the relation
\be\label{e62}
\tr \gamma^\mu \gamma^\nu \gamma^\alpha \gamma^\beta \gamma^5 = - 4 i \epsilon^{\mu\nu\alpha\beta}. 
\ee
Thus, since the trace of $\gamma^5$ with an odd number of $\gamma$-matrices vanishes, we get the relation
\be\label{e63}
\tr\gamma^\nu(\slashed k + m)\gamma^\mu(\slashed k + \slashed p + m)\gamma^5(\slashed k -\slashed q + m)
=4 i m \epsilon^{\mu\nu\alpha\beta} p_\alpha q_\beta 
\ee
which is independent of $k$.
This transverse result % is $k$-independent and
arises in consequence of gauge invariance.

In this way, one can write Eq. \eqref{e61} in the alternative form
\be\label{e64}
\Gamma^{\mu\nu} = -8 g_\pi e^2 \epsilon^{\mu\nu\alpha\beta} p_\alpha q_\beta \, i \int\frac{d^4 k}{(2\pi)^4}
\frac{1}{k^2-m^2+i\epsilon} \frac{1}{(k+p)^2-m^2+i\epsilon} \frac{1}{(k-q)^2-m^2+i\epsilon}.
\ee
We remark that the above integral corresponds to a comtribution coming from a one-loop 3-point scalar function. This can be computed in our approach by using Eq. \eqref{e1} for the scalar propagators, together with the procedure outlined in Sec. \ref{sec2}. After s straightforward calculation, we obtain the result
\be\label{e65}
\Gamma^{\mu\nu} = - 8 g_\pi e^2 \epsilon^{\mu\nu\alpha\beta} p_\alpha q_\beta \Gamma(p,q),
\ee
where the cubic scalar vertex can be written as
\begin{eqnarray}\label{e66}
\Gamma(p,q) = \int\frac{d^3 k}{(2\pi)^3} \frac{1}{2\omega_k} \frac{1}{2\omega_{k+p}} \frac{1}{2\omega_{k-q}}
\left[\frac{1}{\omega_k+\omega_{k+p}-p^0-i\epsilon} \frac{1}{\omega_k+\omega_{k-q}-q^0-i\epsilon} +
\right. \nonumber \\ \left.
\frac{1}{\omega_{k+p}+\omega_{k-q}-p^0-q^0-i\epsilon}\left(\frac{1}{\omega_{k}+\omega_{k+p}-p^0-i\epsilon}
+\frac{1}{\omega_{k}+\omega_{k-q}-q^0-i\epsilon}\right)
+ \left({p^0\rightarrow -p^0}\atop{q^0\rightarrow -q^0}\right)
\right].
\end{eqnarray}
One can see that this contribution has a similar form to that of the last term in Eq. \eqref{e36}.

We will now evaluate the imaginary part of this expression in the physical case when $p^2=0$, $q^2=0$ with $p^0$ and $q^0$ positive. Then, one finds that
\be\label{e67}
\Im \Gamma(p,q) = \pi \int\frac{d^3 k}{(2\pi)^3}\frac{1}{2\omega_k}\frac{1}{2\omega_{k+p}}\frac{1}{2\omega_{k-q}}
\delta(\omega_{k+p}+\omega_{k-q}-p^0-q^0)\left(
\frac{1}{\omega_k+\omega_{k+p}-p^0}+\frac{1}{\omega_k+\omega_{k-q}-q^0}
\right).
\ee
The $\delta$-function corresponds to the contribution coming from the intermediate state involving the cut lines shown in Fig. \ref{fig10n}. This integral can be evaluated with the help of the relation \eqref{e22}, by employing a procedure similar to that used in Eqs. \eqref{e40}--\eqref{e44}. We then get, after some calculation, the Lorentz invariant expression
\be\label{e68}
\Im\Gamma(s) = \frac{1}{16\pi} \frac{1}{s}\ln\left(
\frac{1+\sqrt{1-4 m^2/s}}{1-\sqrt{1-4 m^2/s}} 
\right) \theta(s-4 m^2),
\ee
where $s=(p+q)^2$. Using this imaginary part in the dispersion relation
\be\label{e69}
\Gamma((p+q)^2 ) = \frac{1}{\pi} \int_{4 m^2}^\infty d\sigma \frac{\Im\Gamma(\sigma)}{\sigma-(p+q)^2-i\epsilon}.
\ee
one can compute the full cubic scalar vertex $\Gamma((p+q)^2)$.  Integrating over $\sigma$ and using the Eq. \eqref{e65}, we obtain for the one-loop amplitude $\Gamma^{\mu\nu}(p,q)$ the result
\be\label{e610}
\Gamma^{\mu\nu} = \frac{1}{2\pi^2} g_\pi e^2 \frac{m}{m_\pi^2} \epsilon^{\mu\nu\alpha\beta}p_\alpha q_\beta \int_0^1\frac{dx}{x} \ln\left[1-x(1-x)\frac{m_\pi^2}{m^2-i\epsilon}\right],
  \ee
where we have now set $(p+q)^2 = m_\pi^2$, since the pion is on-shell.

The above result is in agreement with the one found by using the covariant Feynman rules \cite{PeskinSchroeder,weinberg2005quantum,Schwartz2013}. It has been shown that such a result implies that the classical chiral symmetry, which turns up for massless fermions, is violated by quantum effects.

\section{Discussion}\label{sec7}
The usual approach to evaluating Feynman diagrams through the use of the Feynman propagator $\Delta_F(x-y)$ of Eq. \eqref{e1} has the advantage of retaining manifest covariance, while providing the physical interpretation of having a particle propagating forward in time from $y$ to $x$ and an antiparticle propagating similarly from $x$ to $y$.
% positive energy propagating forward in time and negative energy propagating backward in time.
Nevertheless, re-expressing Feynman diagrams in terms of the on-shell propagator $\Delta^+[\pm(x-y)]$ of Eq. \eqref{e2} may be of interest.

Establishing the connection between the perturbative method introduced in Ref. \cite{Kadyshevsky:1967rs} and the usual Feynman approach is our first result. This method makes the determination of the imaginary part of a Feynman diagram quite straightforward, coming as it does entirely from the step function $\theta$. It is somewhat simpler than using the ``largest time'' approach of Ref. \cite{tHooft:1973wagVeltman:1994wz}. Once the imaginary part of the scattering amplitude is determined, a dispersion relation can be used to compute the full amplitude, as was done in Eqs. \eqref{e46}, \eqref{e48}, \eqref{510} and \eqref{e610}.
This alternative to the direct evaluation of Feynman integrals is considered, for example, in Ref. \cite{Elvang:2015rqa}.

As we have shown in scalar models and in QED, the integration over the $\tau$ parameters leads to results that are equivalent to those arising when using the old perturbation theory (OPT).
%, as in Eq. \eqref{e37}.
%Although this approach is not manifestly covariant, it does on occasion provide physical insight (for example, Ref. \cite{Isgur:1972qs}).
Since the present method is based on the forms \eqref{e1} and \eqref{51} which involve the Lorentz invariant Feynman propagator \eqref{e4}, it may be thought of as a covariant formulation of the OPT.
An interesting feature of this method is that it consistently preserves the gauge invariance in QED.
%, as we saw in Eq. \eqref{58}.
In this context, we have examined the vacuum polarization and the decay process of the neutral pion in two photons.

Such an approach is convenient in some circumstances when it provides physical insights on quantum field theory \cite{weinberg2005quantum,Schwartz2013,Isgur:1972qs}. For example,
an appropriate version of OPT is given by Schwinger's proper time formalism, which is the best way to carry out certain effective action calculations \cite{Schwinger:1951nm}.
Moreover, the OPT is also convenient to give a general proof of the infrared finiteness of averaged transition probabilities \cite{GSterman}.
Another helpful feature of OPT is that it clarifies the way that the singularities of the $S$--matrix arise from various physical intermediate states. %\cite{schweber2011introductionweinberg2005quantum}.
This may be useful in connection with the unitarity methods applied to loop computations in gauge theories \cite{bern:1994zxbern:1995cg,Britto:2004nc,Anastasiou:2006gt,Forde:2007mi,Bern:2007dw}, which rely on the fact that loop amplitudes are determined by their singularities.

\begin{acknowledgments}
  {We would like to thank CNPq (Brazil) for financial support.}
%and to J. C. Taylor for a valuable correspondence.
  % D. G. C. M. acknowledges ...
%F. T. B. and  J. F.  thank CNPq (Brazil) for financial support. 
%F. T. B.,  J. F.,  S. M.-F. and G. S. S. S.  thank CNPq (Brazil) for financial support.}
%This work comes as an aftermath of an original 
%project developed with the support of FAPESP (Brazil),  grant number 2018/01073-5.
%D.~G.~C.~M. would like to thank Roger Macleod for helpful discussions}
\end{acknowledgments}

\newpage

\appendix

\section{Unitarity and cut diagrams}\label{appA}
Let us illustrate the equivalence between the imaginary parts of Feynman graphs found in the present approach and the cutting rules of Cutkosky. To this end we consider, for definiteness, the two loop self-energy graph discussed in Sec. \ref{sec2}. As we have shown, performing the $\tau$-integrations, we obtain the same result as that found in the old time-ordered perturbation theory. Thus, the contribution from this graph may be written in the compact form (compare with Eq. \eqref{e36} where we set the interaction vertex equal to $g$)
\be\label{a1}
G(p) = g^4 \int 
\frac{d\gamma_1d\gamma_2d\gamma_3}{(p^0-E_{\gamma_1}+i\epsilon)(p^0-E_{\gamma_2}+i\epsilon)(p^0-E_{\gamma_3}+i\epsilon)},
%+ (p_0 \rightarrow -p_0) ,
\ee
where the integrals over $\gamma_i$ denote integrations over the internal momentum with the corresponding energy factors $1/2\omega_j$, as well as a summation over all time-ordered configurations. Here $E_{\gamma_i}$ is the energy of the intermediate state $\gamma_i$ which contains a sum of mass-shell energies $\omega_j$ of the lines in the state $\gamma_i$ (see, for example, Eq. \eqref{e216})

\begin{figure}[h!!]
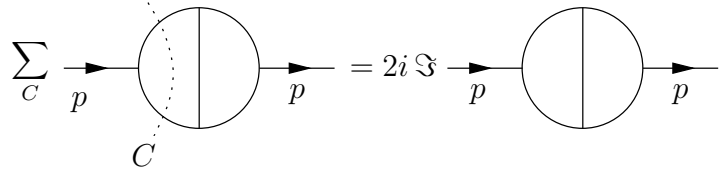

\input fig10.pspdftex
  \caption{Unitarity applied to the two-loop self-energy graph.}
\label{fig10}
\end{figure}

The unitarity condition may be represented as shown in Fig. \ref{fig10}, where the sum is over all cuts $C$.

\noindent
The above relation can be written in the form
\be\label{a2a}
\sum_{\mbox{all } C} G_C(p) = 2 i \Im G(p),
\ee
where the sum over all cuts of $G(p)$ is connected with the decay rate of the particle with momentum $p$.

Let us denote by $G_n$ the graph to the left of the cut involving the on-shell state $\gamma_n$ which has an energy denominator replaced by $2\pi\delta(p^0-E_{\gamma_n})$. The corresponding graph to the right side of the cut, denoted by $G^\star_n$, is computed according to the complex conjugate rules.
Then, using expressions like the one given in Eq. \eqref{a1} and suppressing for simplicity the integrals  and overall factors, we obtain the relation
\be\label{a2}
\sum_n G_n^\star G_n  = \sum_{n=1}^3\left[\prod_{j=n+1}^{3}\frac{1}{p^0-E_{\gamma_j}-i\epsilon}2 \pi\delta(p^0-E_{\gamma_n})\prod_{i=1}^{n-1}\frac{1}{p^0-E_{\gamma_i}+i\epsilon}\right].
\ee
%where contributions with $p_0\rightarrow -p_0$ are to be understood.

On the other hand (omitting the same factors), the imaginary part of $G$ (times $2 i$) is
\be\label{a3}
2 i \,\Im G = i\,\left[\prod_{j=1}^3\frac{1}{p^0-E_{\gamma_j}+i\epsilon}
                      -\prod_{j=1}^3\frac{1}{p^0-E_{\gamma_j}-i\epsilon}
\right].
\ee
The expressions \eqref{a2} and \eqref{a3} are equal, as may be verified using the identity \eqref{e19} which implies that
%fact that terms involving products of $\delta$-functions vanish due to kinematical constraints, together with the
%identity %sssssssss(${\cal P}$ denotes the principal value)
\be\label{a4}
i \left(\frac{1}{A+i\epsilon} - \frac{1}{A-i\epsilon} \right) = 2 \pi \delta(A).
\ee

The equality of the Eqs. \eqref{a2} and \eqref{a3} is equivalent to the unitarity relation \eqref{a2a}. The above derivation may be generalized in a straightforward way to an arbitrary Feynman diagram.
%with $N$ external lines and $V+1$ vertices
(see also the section 9.6 of \cite{GSterman}).
\newpage
%\bibliography{all_new}      

\begin{thebibliography}{10}
% 1
\bibitem{PeskinSchroeder} 
M.~E. Peskin and D. V. Schroeder,
{\em An Introduction to Quantum Field Theory} (Addison-Wesley, Reading, USA, 1995). 
%
% \bibitem{schweber2011introduction} %weinberg2005quantum}
%  See, for example:\\
%  S.~S. Schweber, {\em An Introduction to Relativistic Quantum Field Theory}
%  (Dover Publications, Mineola, NY, 2011).
\bibitem{weinberg2005quantum} 
S. Weinberg, {\em The Quantum Theory of Fields: Volumes I, II and III}
  (Cambridge University Press, Cambridge, UK, 2005). 
\bibitem{Schwartz2013}
M. Schwartz, {\em Quantum Field Theory and the Standard Model} (Cambridge University Press, Cambridge, UK, 2013). %doi:10.1017/9781139540940
% 2
\bibitem{heitler1984quantum}
W. Heitler, {\em The Quantum Theory of Radiation}, {\em Dover Books on Physics}
  (Dover Publications, New York, 1984).
%3
\bibitem{feynman1962quantum}
R. Feynman, {\em Quantum Electrodynamics}, {\em A lecture note and reprint
  volume} (Benjamin, Reading, Massachusetts, 1962).
%4
\bibitem{kaiser2009drawing}
D. Kaiser, {\em Drawing Theories Apart: The Dispersion of Feynman Diagrams in
  Postwar Physics} (The University of Chicago Press, Chicago and London, 2009).
%5
\bibitem{Stueckelberg:1941rg}
E.~C.~G. Stueckelberg, Helv. Phys. Acta {\bf 14}, 588 (1941).%  588.%  (1941).

%\cite{tHooft:1978jhc}
\bibitem{tHooft:1978jhc}
G.~'t Hooft and M.~J.~G.~Veltman,
%``Scalar One Loop Integrals,''
Nucl. Phys. B \textbf{153}, 365 (1979). %, 365. %-401.
%doi:10.1016/0550-3213(79)90605-9
%1510 citations counted in INSPIRE as of 11 May 2022

%6
\bibitem{pauli2000selected}
W. Pauli and C. Enz, {\em Selected Topics in Field Quantization}, {\em Dover
  books on physics} (Dover Publications, Mineola, NY, 2000).
%7
\bibitem{Kadyshevsky:1967rs}
V.~G. Kadyshevsky, Nucl. Phys. B {\bf 6}, 125 (1968). %  125.%  (1968).

%\cite{Fuda:1986zg}
\bibitem{Fuda:1986zg}
M.~G.~Fuda,
%``COVARIANT TIME ORDERED PERTURBATION THEORY,''
Phys. Rev. C \textbf{33}, 996-1001 (1986).
%doi:10.1103/PhysRevC.33.996
%11 citations counted in INSPIRE as of 28 Jun 2022

%8
\bibitem{Cutkosky:1960sp}
R.~E. Cutkosky, J. Math. Phys. {\bf 1},  (1960)  429.%  (1960).
%9
\bibitem{Elvang:2015rqa}
H. Elvang and Y.-t. Huang, {\em {Scattering Amplitudes in Gauge Theory and
  Gravity}} (Cambridge University Press, Cambridge, UK, 2015);\\
%\bibitem{Gribov:2000nhz}
V.~N. Gribov and J. Nyiri, {\em {Quantum electrodynamics: Gribov lectures on
  theoretical physics}} (Cambridge University Press, Cambridge, UK, 2005),
  Vol.~13;\\
%\bibitem{Berestetskii:1982qgu}
V.~B. Berestetskii, E.~M. Lifshitz, and L.~P. Pitaevskii, {\em {Quantum
  Electrodynamics}}, Vol.~4 of {\em Course of Theoretical Physics} (Pergamon
  Press, Oxford, 1982);\\ 
%\bibitem{Broadhurst:1987ei}
D.~J. Broadhurst, Z. Phys. C {\bf 47}, 115 (1990); \\ %  115; \\ %  (1990);\\
%\bibitem{vanNeerven:1985xr}
W.~L. van Neerven, Nucl. Phys. B {\bf 268}, 463 (1986). %  453.%  (1986).
%10

%11
\bibitem{Feynman:1963axKlauder:1972lsv}
R.~P. Feynman, Acta Phys. Polon. {\bf 24}, 697 (1963); %  697; %  (1963);
%\bibitem{Klauder:1972lsv}
{\em {Magic without magic: John Archibald Wheeler}: {A collection of essays in
  honor of his sixtieth birthday}}, edited by J.~R. Klauder (Freeman, San
  Francisco, 1972).
  % 12

\bibitem{Berger:2009zb}
C.~F. Berger and D. Forde, Ann. Rev. Nucl. Part. Sci. {\bf 60}, 181 (2010). %  181. %  (2010).


\bibitem{Brandhuber:2005kd}
A. Brandhuber, B. Spence, and G. Travaglini, JHEP {\bf 01}, 142 (2006). %  142. %  (2006).

\bibitem{Rodrigo:2008fp}
G. Rodrigo, S. Catani, T. Gleisberg, F. Krauss, and J.-C. Winter, Nucl. Phys. B
  Proc. Suppl. {\bf 183}, 262 (2008). %  262. %  (2008).

\bibitem{Bierenbaum:2010cy}
I. Bierenbaum, S. Catani, P. Draggiotis, and G. Rodrigo, JHEP {\bf 10}, 073 (2010). %  073. %  (2010).

\bibitem{Caron-Huot:2010fvq}
S. Caron-Huot, JHEP {\bf 05}, 080 (2011). %  080. %  (2011).

%\cite{Brandt:2021nse}
\bibitem{Brandt:2021nseBrandt:2021nev}
F.~T.~Brandt, J.~Frenkel, S.~Martins-Filho, D.~G.~C.~McKeon and G.~S.~S.~Sakoda,
%``Forward scattering amplitudes in the imaginary time formalism,''
Phys. Rev. D \textbf{104}, no.10, 105007 (2021); % 105007; % (2021)
%doi:10.1103/PhysRevD.104.105007
%[arXiv:2110.07694 [hep-th]].
%0 citations counted in INSPIRE as of 12 Apr 2022
%\bibitem{Brandt:2021nev}
%F.~T.~Brandt, J.~Frenkel, S.~Martins-Filho, D.~G.~C.~McKeon and G.~S.~S.~Sakoda,
%``Thermal gauge theories with Lagrange multiplier fields,''
Can. J. Phys. \textbf{100}, no. 3, 139 (2022). % no.3, 139. %-144.
%doi:10.1139/cjp-2021-0248
%[arXiv:2105.00318 [hep-th]].
%0 citations counted in INSPIRE as of 09 May 2022


\bibitem{tHooft:1972fiBollini:1972ui}
G. 't~Hooft and M.~J.~G. Veltman, Nucl. Phys. {\bf B44}, 189 (1972); %  189; %  (1972);
%\bibitem{Bollini:1972ui}
C.~G. Bollini and J.~J. Giambiagi, Nuovo Cim. {\bf B12}, 20 (1972). %  20. %  (1972).
%13
\bibitem{tHooft:1973wagVeltman:1994wz}
G. 't~Hooft and M.~J.~G. Veltman, NATO Sci. Ser. B {\bf 4}, 177 (1974); %  177; %  (1974);
%\bibitem{Veltman:1994wz}
M.~J.~G. Veltman, {\em {Diagrammatica: The Path to Feynman rules}} (Cambridge
University Press, Cambridge, UK, 2012), Vol.~4.

%\cite{Delbourgo:2003zi}
\bibitem{Delbourgo:2003ziDavydychev:2003cw}
R.~Delbourgo and M.~L.~Roberts,
%``Relativistic phase space: Dimensional recurrences,''
J. Phys. A \textbf{36}, 1719-1728 (2003); \\
%doi:10.1088/0305-4470/36/6/315
%arXiv:hep-th/0301004 [hep-th]].
%5 citations counted in INSPIRE as of 28 Jun 2022
%\cite{Davydychev:2003cw}
%\bibitem{Davydychev:2003cw}
A.~I.~Davydychev and R.~Delbourgo,
%``Explicitly symmetrical treatment of three body phase space,''
J. Phys. A \textbf{37}, 4871-4886 (2004).
%doi:10.1088/0305-4470/37/17/016
%[arXiv:hep-th/0311075 [hep-th]].
%20 citations counted in INSPIRE as of 28 Jun 2022

\bibitem{grads}
I. S. Gradshteyn and I. M. Ryzhik,
  {\em Table of Integrals, Series, and Products} (Academic Press, London, UK, 1980).
  
\bibitem{AdlerBellJackiw}
S. L. Adler, Phys. Rev. {\bf 177}, 2426 (1969); J. S. Bell and R. Jackiw, Nuovo Cimento {\bf 60A}, 47 (1969).
  
% 14a
%\cite{Brandt:2021nse}
%\bibitem{Brandt:2021nse}
%F.~T.~Brandt, J.~Frenkel, S.~Martins-Filho, D.~G.~C.~McKeon and G.~S.~S.~Sakoda,
%``Forward scattering amplitudes in the imaginary time formalism,''
%Phys. Rev. D \textbf{104}, no.10, 105007 (2021).
%doi:10.1103/PhysRevD.104.105007
%[arXiv:2110.07694 [hep-th]].
%0 citations counted in INSPIRE as of 01 Apr 2022
% 14
\bibitem{Isgur:1972qs}
N. Isgur, Phys. Rev. D {\bf 6}, 393 (1972). %  393. %  (1972).

%\cite{Schwinger:1951nm}
\bibitem{Schwinger:1951nm}
J.~S.~Schwinger,
%``On gauge invariance and vacuum polarization,''
Phys. Rev. \textbf{82}, 664 (1951). %, 664. %-679
%doi:10.1103/PhysRev.82.664
%5573 citations counted in INSPIRE as of 18 May 2022

%\bibitem{Feynman:1963ax}
%R.~P. Feynman, Acta Phys. Polon. {\bf 24},  697  (1963).

\bibitem{GSterman}
G. Sterman,
{\em An Introduction to Quantum Field Theory} (Cambridge University Press, Cambridge, UK, 1993).

\bibitem{bern:1994zxbern:1995cg}
 Z. Bern, L. Dixon, D.~C. Dunbar, and D.~A. Kosower, Nucl. Phys. {\bf B425}, 271 (1994); % 217; %  (1994);
%
%\bibitem{bern:1995cg}
%Z. Bern, L. Dixon, D.~C. Dunbar, and D.~A. Kosower,
Nucl. Phys. {\bf B435}, 59 (1995). %  59. %   (1995).

\bibitem{Britto:2004nc}
R. Britto, F. Cachazo, and B. Feng, Nucl. Phys. B {\bf 725}, 275 (2005). %  275. %  (2005).

\bibitem{Anastasiou:2006gt}
C. Anastasiou, R. Britto, B. Feng, Z. Kunszt, and P. Mastrolia, JHEP {\bf 03}, 111 (2007). % 111. %  (2007).
  
\bibitem{Forde:2007mi}
D. Forde, Phys. Rev. D {\bf 75}, 125019 (2007). %  125019. %  (2007).

\bibitem{Bern:2007dw}
Z. Bern, L.~J. Dixon, and D.~A. Kosower, Annals Phys. {\bf 322}, 1587 (2007). %  1587. %  (2007).

\end{thebibliography}
%\bibliographystyle{prsty}

% American style

\end{document}